\documentclass[
preprint,
superscriptaddress,
 amsmath,amssymb,
 aps, physrev,
floatfix,
pre,
]{revtex4-1}

\usepackage{graphicx}
\usepackage{dcolumn}
\usepackage{bm}

\begin{document}

\title{Response of social norms to individual differences in error-proneness}%

\author{Quang Anh Le}
\affiliation{Industry-University Cooperation Foundation, Pukyong National
University, Busan 48513, South Korea}
\author{Seung Ki Baek}
\email{seungki@pknu.ac.kr}
\affiliation{Department of Scientific Computing, Pukyong National University,
Busan 48513, South Korea}
\date{\today}

\begin{abstract}
Indirect reciprocity explains the evolution of cooperation by considering how our cooperative behavior toward someone is reciprocated by someone else who has observed us. A cohesive society has a shared norm that prescribes how to assess observed behavior as well as how to behave toward others based on the assessments, and the eight social norms that are evolutionarily stable against the invasion of mutants with different behavioral rules are referred to as the leading eight, whose member norms are called L1 to L8, respectively.
Among the leading eight, L8 (also known as `Judging') has been deemed mostly irrelevant due to its poor performance in maintaining cooperation when each person may have a different opinion about someone instead of forming a public consensus. In this work, we propose that L8 can nevertheless be best protected from assessment errors among the leading eight if we take into account the intrinsic heterogeneity of error proneness among individuals because this norm heavily punishes those who are prone to errors in following its assessment rule. This finding suggests that individual differences should be explicitly taken into account as quenched randomness to obtain a thorough understanding of a social norm working in a heterogeneous environment.
\end{abstract}

\maketitle


\section{Introduction}

Indirect reciprocity is one of the main mechanisms for promoting the evolution of cooperation through the interaction of assessment and behavior~\cite{nowak1998evolution,nowak2005evolution,ohtsuki2004should}.
The conditions for a social norm to maintain cooperation against external perturbations have been revealed since the discovery of the `leading eight' (Table~\ref{tab:eight})~\cite{ohtsuki2006leading,ohtsuki2009indirect}.
Each norm in the leading eight achieves stable cooperation against the invasion of mutant norms with different behavioral rules, as long as the society has a public consensus in assessing each individual.
This line of research has been extended to a more realistic situation in which assessments are made privately, instead of forming a consensus, and it has turned out that privateness renders many of the leading eight untenable~\cite{hilbe2018indirect,murase2024computational}.
In particular, we have seen that strict norms such as Judging and Stern Judging (also known as L8 and L6, respectively) fail to sustain cooperation when disagreement arises~\cite{lee2022second,bae2024exact}.
For example, Judging divides an all-to-all connected society into a weakly balanced configuration~\cite{easley2010networks} with many antagonistic groups~\cite{bae2025indirect}, resulting in the lowest level of cooperation among the leading eight~\cite{fujimoto2024leader}. However, weakly balanced structures are often found in empirical social networks~\cite{leskovec2010signed,szell2010multirelational}, suggesting a possible selective advantage of Judging. In this work, we wish to explain why Judging is a special norm in terms of robustness in a noisy environment.

\begin{table}
\caption{\label{tab:eight}The leading eight. Each of the eight norms has an assessment rule $\alpha$ and a behavioral rule $\beta$. An observer is observing an interaction between a donor and a recipient, where the donor chooses behavior between cooperation (C) and defection (D), and the observer regards the donor's behavior as either good (G) or bad (B).
The assessment rule tells the observer to assign $\alpha_{XYZ}\in\{\text{G},\text{B}\}$ to the donor when the observer regards the donor as $X\in\{\text{G},\text{B}\}$, the donor does $Y\in\{\text{C},\text{D}\}$ to the recipient, and the observer regards the recipient as $Z\in\{\text{G},\text{B}\}$. The donor chooses behavior $\beta_{XY} \in \{\text{C},\text{D}\}$ to the recipient when the donor's self-assessment is $X\in\{\text{G},\text{B}\}$ and the donor regards the recipient as $Y\in\{\text{G},\text{B}\}$.}
\begin{ruledtabular}
\begin{tabular}{c|cccccccc|cccc}
& $\alpha_{\text{GCG}}$ & $\alpha_{\text{GDG}}$ & $\alpha_{\text{GCB}}$ & $\alpha_{\text{GDB}}$ &
 $\alpha_{\text{BCG}}$ & $\alpha_{\text{BDG}}$ & $\alpha_{\text{BCB}}$ & $\alpha_{\text{BDB}}$ &
 $\beta_{\text{GG}}$ & $\beta_{\text{GB}}$ & $\beta_{\text{BG}}$ & $\beta_{\text{BB}}$\\\hline
L1 & G & B & G & G & G & B & G & B & C & D & C & C\\
L2 (Consistent Standing) & G & B & B & G & G & B & G & B & C & D & C & C\\
L3 (Simple Standing) & G & B & G & G & G & B & G & G & C & D & C & D\\
L4 & G & B & G & G & G & B & B & G & C & D & C & D\\
L5 & G & B & B & G & G & B & G & G & C & D & C & D\\
L6 (Stern Judging) & G & B & B & G & G & B & B & G & C & D & C & D\\
L7 (Staying) & G & B & G & G & G & B & B & B & C & D & C & D\\
L8 (Judging) & $G$ & $B$ & $B$ & $G$ & $G$ & $B$ & $B$ & $B$ & C & D & C & D\\
\end{tabular}
\end{ruledtabular}
\end{table}

From a methodological point of view, the assumption of private assessment can be viewed as an attempt to go beyond a mean-field approach, which has usually been used by considering a well-mixed population (see, however, Ref.~\cite{murase2024computational} demonstrating the importance of population structure in indirect reciprocity).
The mean-field approximation greatly reduces the number of degrees of freedom to make the problem tractable, but the price is that it loses every piece of information about individual differences.
As a way of retaining individual differences, this work assumes that each individual has a different probability of error. Some are hasty in assessing others, while some others are more prudent and seldom make mistakes in recognizing someone's goodness. The existence of such individual differences is obvious, and it sounds plausible that erroneous defection will be harmful to one's own reputation in a cooperative society. However, to our knowledge, the correlation between an individual's error probability and his or her overall reputation in the long run has not yet been investigated.
In this work, we will present some analytic progresses on this issue.

Before proceeding, let us classify errors into three types. The first is an assessment error, which means that an observer remembers a donor as good, although the correct assessment should be the opposite, or vice versa. 
This is the type of error that we will focus on throughout this work.
The second is a perception error, by which an observer mistakes a donor's cooperation as defection, or vice versa. Note the difference between the assessment error and the perception error: If the observer is an unconditional cooperator, the perception error does not change the observer's assessment, whereas the assessment error does. However, if we work with the leading eight in the vicinity of paradise where everyone is good, the assessment is heavily based on the observed behavior, so the perception error plays a similar role to that of the assessment error.
The last is a behavioral error. A donor cooperates by error, although the correct behavior is defection, or vice versa. Later, we will examine the effects of behavioral errors through numerical calculations.

\section{Analysis}

We consider a population of size $N \gg 1$.
Let $m_{ij}^t$ denote how an individual $i$ assesses an individual $j$ at time $t$. If $i$ regards $j$ as perfectly good (bad), we have $m_{ij}^t=1 (0)$, but it is generally between zero and one~\cite{lee2021local,lee2022second,mun2023second,mun2024making}. The conventional discrete model can be said to use only the end points.
The system becomes more analytically tractable when we work with continuous variables. Even if we consider the discrete model, we expect that the continuous description can capture the average behavior involved with probabilistic errors. To see the correspondence between these two approaches, we will carry out analytic calculations within the continuous model, while the discrete model is used in numerical simulations.

For each round, a pair of randomly chosen individuals $i$ and $j$ interact with each other by playing the donation game, where $i$ as a donor can choose to cooperate or defect. If the donor cooperates, the donor's payoff decreases by $c$ as the cost of cooperation, and the other individual $j$ playing the role of the recipient earns $b$ as the benefit of cooperation. However, if the donor defects, it means that the donor refuses to cooperate and their payoffs do not change. When $b>c>0$, the donation game is a special type of prisoner's dilemma.
What $i$ does to $j$ is determined by his or her behavior rule $\beta_i$, which depends on $i$'s self-assessment as well as on how $i$ regards $j$. Mathematically speaking, this can be expressed by $\beta_i = \beta_i \left( m_{ii}^t, m_{ij}^t \right)$.
In the continuous version, the donor $i$ pays the cost of cooperation $c\beta_i$, which benefits the recipient by $b\beta_i$, so that $\beta_i=1$ and $0$ mean full cooperation and defection, respectively.
Every observer $k$ observes the interaction between $i$ and $j$ with probability $q$ and assesses the donor $i$ according to his or her own assessment rule $\alpha_k$.
The assessment depends on how $k$ regards $i$, what $i$ does to $j$, and how $k$ regards $j$, which can be expressed by $\alpha_k = \alpha_k \left[ m_{ki}^t, \beta_i \left( m_{ii}^t, m_{ij}^t \right), m_{kj} ^t\right]$. However, with probability $\sigma_{ki}$, the assessment can be flipped to $1-\alpha_k$.
An individual $k$'s social norm is the combination of the assessment rule $\alpha_k$ and the behavioral rule $\beta_k$. Table~\ref{tab:cont} shows the continuous versions of the leading eight obtained through bilinear and trilinear interpolations so that the original definitions are recovered at the end points.

\begin{table}
\caption{\label{tab:cont}Continuous expressions of the leading eight obtained through bi- and tri-linear interpolations. The last column shows the value of $\alpha$ when $m_{ki}=1/2$ for every pair of $k$ and $i$.}
\begin{ruledtabular}
\begin{tabular}{lccc}
        Norm & $\alpha(x,y,z)$ & $\beta(x,y)$ & $\alpha^\ast \equiv \alpha\left[\frac{1}{2}, \beta\left(\frac{1}{2},\frac{1}{2}\right), \frac{1}{2}\right]$ \\\hline
        L1 & $x+y-xy-xz+xyz$ & $-x+xy+1$ & 13/16\\
        L2 (Consistent Standing) & $x+y-2xy-xz+2xyz$ & $-x+xy+1$ & 5/8\\
        L3 (Simple Standing) & $yz - z + 1$ & $y$ & 3/4\\
        L4 & $-y-z+xy+2yz-xyz+1$ & $y$ & 5/8\\
        L5 & $-z-xy+yz+xyz+1$ & $y$ & 5/8\\
        L6 (Stern Judging) & $-y-z+2yz+1$ & $y$ & 1/2\\
        L7 (Staying) & $x-xz+yz$ & $y$ & 1/2\\
        L8 (Judging) & $x-xy-xz+yz+xyz$ & $y$ & 3/8\\
\end{tabular}
\end{ruledtabular}    
\end{table}

The above dynamical rule can be written as the following equation:
\begin{eqnarray}
m_{ki}^{t+1} = (1-q) m_{ki}^t &+& \frac{q}{N} \sum_{j=1}^N (1-\sigma_{ki}) \alpha_k \left[ m_{ki}^t, \beta_i(m_{ii}^t, m_{ij}^t), m_{kj}^t \right]\nonumber\\
&&+ \sigma_{ki} \left\{ 1- \alpha_k \left[ m_{ki}^t, \beta_i(m_{ii}^t, m_{ij}^t), m_{kj}^t \right]  \right\},
\label{eq:dynamics}
\end{eqnarray}
where $\sigma_{ki}$ is the probability of assessment error, which may depend on the observer $k$ or the donor $i$. In the second term on the right-hand side, we have taken the average over the randomly chosen recipient $j$, which may also be equal to $i$ for mathematical convenience.
Assuming that everyone uses the same norm, we can say $\alpha_k = \alpha$ and $\beta_i = \beta$ without the subscripts.
Rearranging the terms of Eq.~\eqref{eq:dynamics} in the long-time limit, we have the following $N^2$-dimensional system of equations to solve:
\begin{equation}
0 = -m_{ki} + \sigma_{ki} + \frac{(1-2\sigma_{ki})}{N} \sum_{j=1}^N \alpha \left[ m_{ki}, \beta(m_{ii}, m_{ij}), m_{kj} \right],    
\label{eq:equation}
\end{equation}
where $\alpha$ and $\beta$ are given in Table~\ref{tab:cont} and the superscripts can now be neglected. Note that the observation probability $q$ becomes irrelevant in this steady state. It is clearly seen that $m_{ki}=1/2$ if $\sigma_{ki}=1/2$, which means that the observer makes random assessments.

\subsection{How an error-prone individual assesses others}

As a specific example, consider L3. We will furthermore assume that the error probability depends only on the observer so that $\sigma_{ki} = \sigma_k$. Equation~\eqref{eq:equation} is then rewritten as
\begin{equation}
0 = -m_{ki} + \sigma_{k} + (1-2\sigma_{k}) \left( C_{ki} - \mu_k + 1 \right),
\label{eq:L3}
\end{equation}
where $C_{ki} \equiv N^{-1} \sum_j m_{kj} m_{ij}$ and $\mu_k \equiv N^{-1} \sum_j m_{kj}$.
By assuming that $(m_{kj}-\mu_k)$ and $(m_{ij}-\mu_i)$ fluctuate independently, we replace $C_{ki}$ by $\mu_k \mu_i$ to get
\begin{equation}
0 \approx -m_{ki} + \sigma_{k} + (1-2\sigma_{k}) \left( \mu_k \mu_i - \mu_k + 1 \right).
\label{eq:mean-field}
\end{equation}
Summing both sides of Eq.~\eqref{eq:L3} over $i$ and dividing them by $N$, we obtain
\begin{equation}
0 \approx -\mu_k + \sigma_{k} + (1-2\sigma_{k}) \left( \mu_k \bar{\mu} - \mu_k + 1 \right),
\label{eq:averages}
\end{equation}
where $\bar{\mu} \equiv N^{-1} \sum_{i} \mu_i$. We postulate that how an observer $k$ assesses others is determined by the observer's probability of error, so that we can write $\mu_k = \mu\left(\sigma_k\right)$. The simplest functional form would be a linear function such as $\mu\left(\sigma_k\right) = u \sigma_k + v$ with constants $u$ and $v$. To satisfy $\mu\left(\sigma_k=1/2\right) = 1/2$, we have $v = (1-u)/2$, which means that
\begin{equation}
\mu_k = \mu \left( \sigma_k \right) = u \sigma_k + \frac{1}{2}(1-u).
\label{eq:mu_k}
\end{equation}
If $\sigma_k$ is uniformly distributed between $0$ and $1/2$, we should have
\begin{equation}
\bar{\mu} = \frac{1}{4} (2-u)
\label{eq:bar_mu}
\end{equation}
in the large-$N$ limit because
\begin{equation}
\lim_{N\to\infty}\frac{1}{N} \sum_{k=1}^N \sigma_k^n = \frac{\int_0^{\frac{1}{2}} x^n dx}{\int_0^{\frac{1}{2}} dx} = \frac{1}{(n+1)2^n}.
\label{eq:sigma_k^n}
\end{equation}
Substituting Eqs.~\eqref{eq:mu_k} and \eqref{eq:bar_mu} into Eq.~\eqref{eq:averages} and averaging both the sides over $k$ by using Eq.~\eqref{eq:sigma_k^n}, we get an algebraic equation for $u$, whose physical solution is $u=-1/2$. Thus, Eq.~\eqref{eq:mu_k} in this case predicts
\begin{equation}
\mu \left(\sigma_k \right) = \frac{1}{4} \left(3-2\sigma_k \right).
\label{eq:ss_mu_k}
\end{equation}
When $k=i$, Eq.~\eqref{eq:mean-field} leads to
\begin{equation}
m_{kk} \approx \sigma_{k} + (1-2\sigma_{k}) \left( \mu_k^2 - \mu_k + 1 \right),
\label{eq:kk}
\end{equation}
by assuming that $C_{kk} - \mu_k^2$, i.e., the variance of $m_{kk}$, vanishes. Even if this assumption cannot be justified in general, suppose that it is valid in the vicinity of $\sigma_k = 1/2$, where $m_{ki}$ is identically equal to $1/2$. Regarding $m_{kk}$ as a proxy of $\mu_k$, we can explicitly solve Eq.~\eqref{eq:kk} for $\mu_k$ and obtain
\begin{equation}
\mu_k = \frac{1-\sigma_k-\sqrt{\sigma_k(1-\sigma_k)}}{1-2\sigma_k} = \frac{1}{4} \left( 3-2\sigma_k \right) + O\left(\left|\frac{1}{2} - \sigma_k \right|^3 \right),
\label{eq:one-body}
\end{equation}
reproducing Eq.~\eqref{eq:ss_mu_k} near $\sigma_k = 1/2$. We stress that replacing $m_{kk}$ by $\mu_k$ is only an approximation to obtain $\mu_k$, which is actually dominated by $m_{ki}$ with $i \neq k$.

\begin{table}
\caption{\label{tab:first}First-order derivatives of $\alpha$ and $\beta$ when $m_{ki}=1/2$ for every pair of $k$ and $i$.}
\begin{ruledtabular}
\begin{tabular}{@{}cccccc@{}}
        Norm & $A'_x$ & $A'_y$ & $A'_z$ & $B'_x$ & $B'_y$ \\\hline
        L1 & 1/8 & 3/4 & -1/8 & -1/2 & 1/2\\
        L2 (Consistent Standing) & -1/4 & 1/2 & 1/4 & -1/2 & 1/2\\
        L3 (Simple Standing) & 0 & 1/2 & -1/2 & 0 & 1\\
        L4 & 1/4 & 1/4 & -1/4 & 0 & 1\\
        L5 & -1/4 & 1/4 & -1/4 & 0 & 1\\
        L6 (Stern  Judging) & 0 & 0 & 0 & 0 & 1\\
        L7 (Staying) & 1/2 & 1/2 & 0 & 0 & 1\\
        L8 (Judging) & 1/4 & 1/4 & 1/4 & 0 & 1\\
\end{tabular}
\end{ruledtabular}
\end{table}

\begin{figure}
\includegraphics[width=0.8\columnwidth]{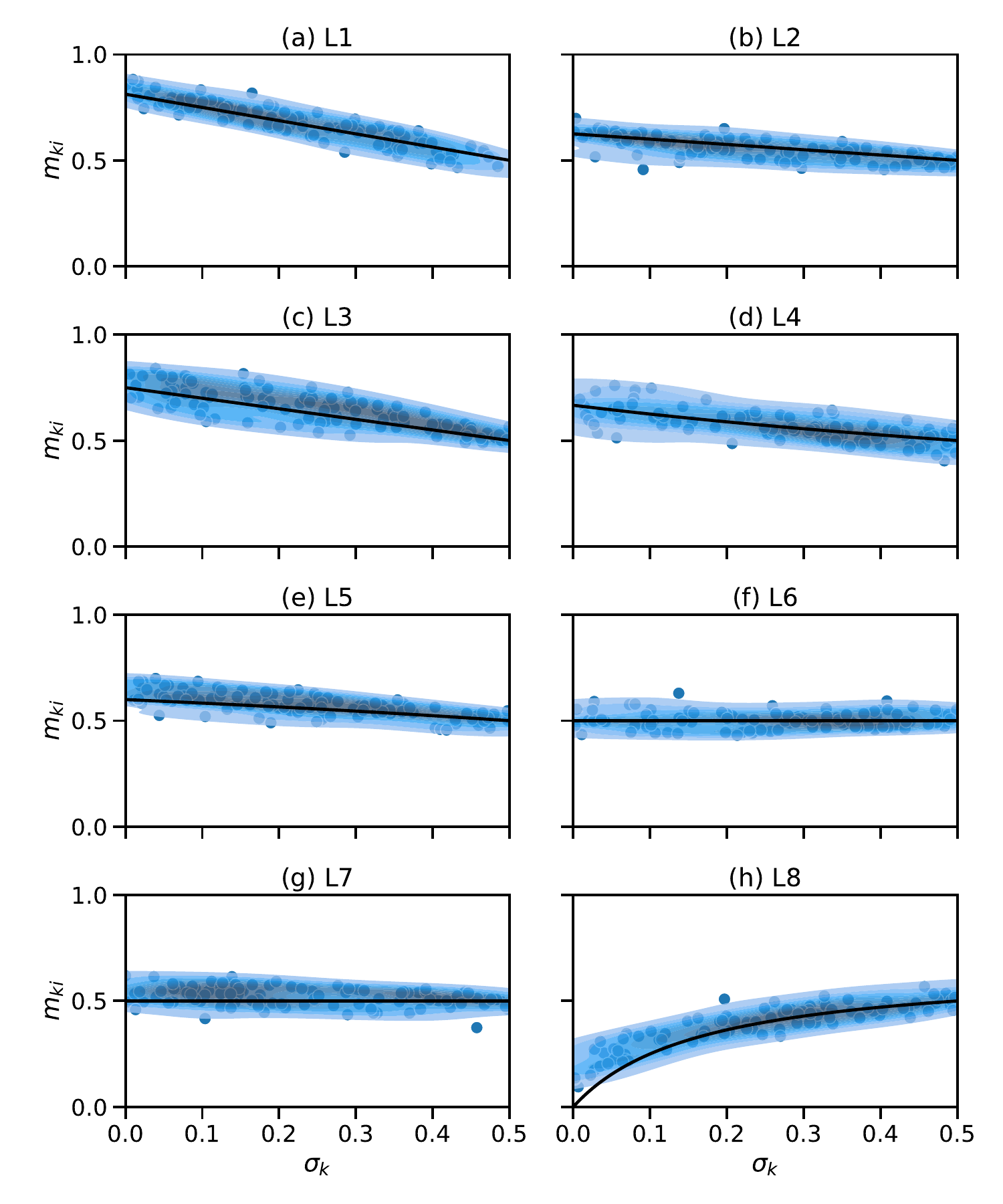}
\caption{Numerical results from the discrete model in which $m_{ki}$ can take only $0$ and $1$. The size of the population is $N=10^2$, and the observation probability is $q=1$. The horizontal axis is the probability of assessment error of an individual $k$, and the vertical axis is the time average of $m_{ki}$ for $T=5\times10^4$ rounds, after discarding transient data for the first $T$ rounds.
In every panel, we have $10^2$ data points, each of which has been obtained by picking up two specific individuals $k$ and $i$ from a different sample. Each sample runs with an independent realization of $\left\{ \sigma_k \right\}$ as a set of quenched random numbers between $0$ and $1/2$.
As the initial condition, we fill the image matrix $\left\{ m_{ki} \right\}$ with random numbers uniformly distributed in the unit interval. The shades show the kernel-density estimates~\cite{Waskom2021}, and the solid lines are obtained from Eq.~\eqref{eq:mkk_general}.
}
\label{fig:out}
\end{figure}

A convenient way to solve such a nonlinear equation is the Newton method~\cite{baek2017duality}.
To see a one-dimensional example for solving $f(x) = 0$, let us denote a trial solution as $\hat{x}$, while the unknown true solution is denoted as $x^\ast$.
We expand the equation around the trial solution to the first order as follows:
\begin{equation}
0 = f \left(x^\ast\right) = f (\hat{x}) + \left(x^\ast-\hat{x}\right) \left.\frac{df}{dx}\right|_{\hat{x}} + \ldots,
\end{equation}
and we observe that the true solution is approximated as
\begin{equation}
x^\ast \approx \hat{x} - \frac{1}{\left( \left. df/dx\right|_{\hat{x}} \right)} f (\hat{x}).
\label{eq:Newton}
\end{equation}
In our problem, the trial solution should be $\hat{\mu}_k=1/2$, which is an exact solution for $\sigma_k=1/2$. To apply the Newton method, we need first-order derivatives evaluated at this trial solution.
Let us rewrite $m_{kk} \approx \mu_k = 1/2 + \delta_k$ and expand Eq.~\eqref{eq:kk} to the first order of $\delta_k$ as follows:
\begin{eqnarray}
0 &=& -\left( \frac{1}{2} + \delta_k \right) + \sigma_k + (1-2\sigma_k) \alpha \left[ \frac{1}{2} + \delta_k, \beta\left(\frac{1}{2} + \delta_k, \frac{1}{2} + \delta_k\right), \frac{1}{2} + \delta_k \right]\\
&\approx& -\left( \frac{1}{2} + \delta_k \right) + \sigma_k + (1-2\sigma_k) \left[ \alpha^\ast 
 + A'_x \delta_k + A'_y \left(B'_x \delta_k + B'_y \delta_k \right) + A'_z \delta_k \right],
\end{eqnarray}
where $\alpha^\ast \equiv \alpha \left[ \frac{1}{2}, \beta\left(\frac{1}{2}, \frac{1}{2} \right), \frac{1}{2} \right]$, $A'_\xi \equiv \left( \partial \alpha / \partial \xi \right)_{(x,y,z)=\left( \frac{1}{2}, \beta\left(\frac{1}{2}, \frac{1}{2} \right), \frac{1}{2} \right)}$, and $B'_\xi \equiv \left( \partial \beta / \partial \xi \right)_{(x,y)=\left( \frac{1}{2}, \frac{1}{2} \right)}$ (see Table~\ref{tab:first}).
Equation~\eqref{eq:Newton} then yields
\begin{equation}
\mu_k^\ast \approx \frac{1}{4} \left( 3-2\sigma_k \right),
\end{equation}
in agreement with Eq.~\eqref{eq:ss_mu_k}. As for the other norms, the general expression is given as follows:
\begin{equation}
\mu_k^\ast \approx \frac{1}{2} \left\{ 1+\frac{(2 \alpha^\ast-1) (1-2 \sigma_k)}{1-(1-2\sigma_k) \left[ A'_x+A'_y (B'_x+B'_y)+A'_z \right]} \right\}.
\label{eq:mkk_general}
\end{equation}
This formula predicts that only L8 will exhibit positive correlation between $\sigma_k$ and $\mu_k$, and the reason is that only L8 has $\alpha^\ast < 1/2$ (Table~\ref{tab:cont}). The correlation vanishes for L6 and L7 because $\alpha^\ast = 1/2$. The behavior of L6 is totally driven by entropy, as has already been analyzed in detail~\cite{bae2024exact}.
Concerning L7, as long as $\alpha^\ast = 1/2$, the steady-state equation [Eq.~\eqref{eq:equation}] actually admits a solution such that $m_{ki}=1/2$ for every pair of $k$ and $i$, regardless of the distribution of $\left\{ \sigma_k \right\}$. For the other five norms from L1 to L5, the correlation is negative, which means that a careless individual tends to assign low assessments to others. Figure~\ref{fig:out} shows that all these predictions are well corroborated by numerical simulations.

\subsection{How an error-prone individual is assessed by others}

\begin{table}
\caption{\label{tab:observer}Ranges of $m_{ki}$ 
 that an observer $i$ receives from others under different social norms, when the probability of error depends on the observer, i.e., $\sigma_{ij} = \sigma_i$. The ranges are obtained by applying the Newton method~\cite{Mathematica}.}
\begin{ruledtabular}
\begin{tabular}{@{}ccc@{}}
 & $N=2$ & $N=3$ \\\hline
L1 & $\frac{1}{2} \le m_{ki} \le \frac{13}{16}$ & $\frac{1}{2} \le m_{ki} \le \frac{13}{16}$\\
L2 (Consistent Standing) & $\frac{1}{2} \le m_{ki} \le \frac{5}{8}$ & $\frac{1}{2} \le m_{ki} \le \frac{5}{8}$\\
L3 (Simple Standing) & $\frac{1}{2} \le m_{ki} \le \frac{27-15\sigma_i}{36-12\sigma_i}$ & $\frac{1}{2} \le m_{ki} \le \frac{27-18\sigma_i}{36-16\sigma_i}$\\
L4 & $\frac{1}{2} \le m_{ki} \le \frac{18+3\sigma_i}{27+9\sigma_i}$ & $\frac{1}{2} \le m_{ki} \le \frac{18+\sigma_i}{27+6\sigma_i}$\\
L5 & $\frac{1}{2} \le m_{ki} \le \frac{135-78\sigma_i}{225-120\sigma_i}$ & $\frac{1}{2} \le m_{ki} \le \frac{135-84\sigma_i}{225-130\sigma_i}$\\
L6 (Stern Judging) & $m_{ki}=\frac{1}{2}$ & $m_{ki}=\frac{1}{2}$\\
L7 (Staying) & $m_{ki}=\frac{1}{2}$ & $m_{ki}=\frac{1}{2}$\\
L8 (Judging) & $\frac{1}{2} \le m_{ki} \le \frac{12\sigma_i}{9+36\sigma_i}$ & $\frac{1}{2} \le m_{ki} \le \frac{10\sigma_i}{9+30\sigma_i}$
\end{tabular}
\end{ruledtabular}
\end{table}

\begin{figure}
\includegraphics[width=0.8\columnwidth]{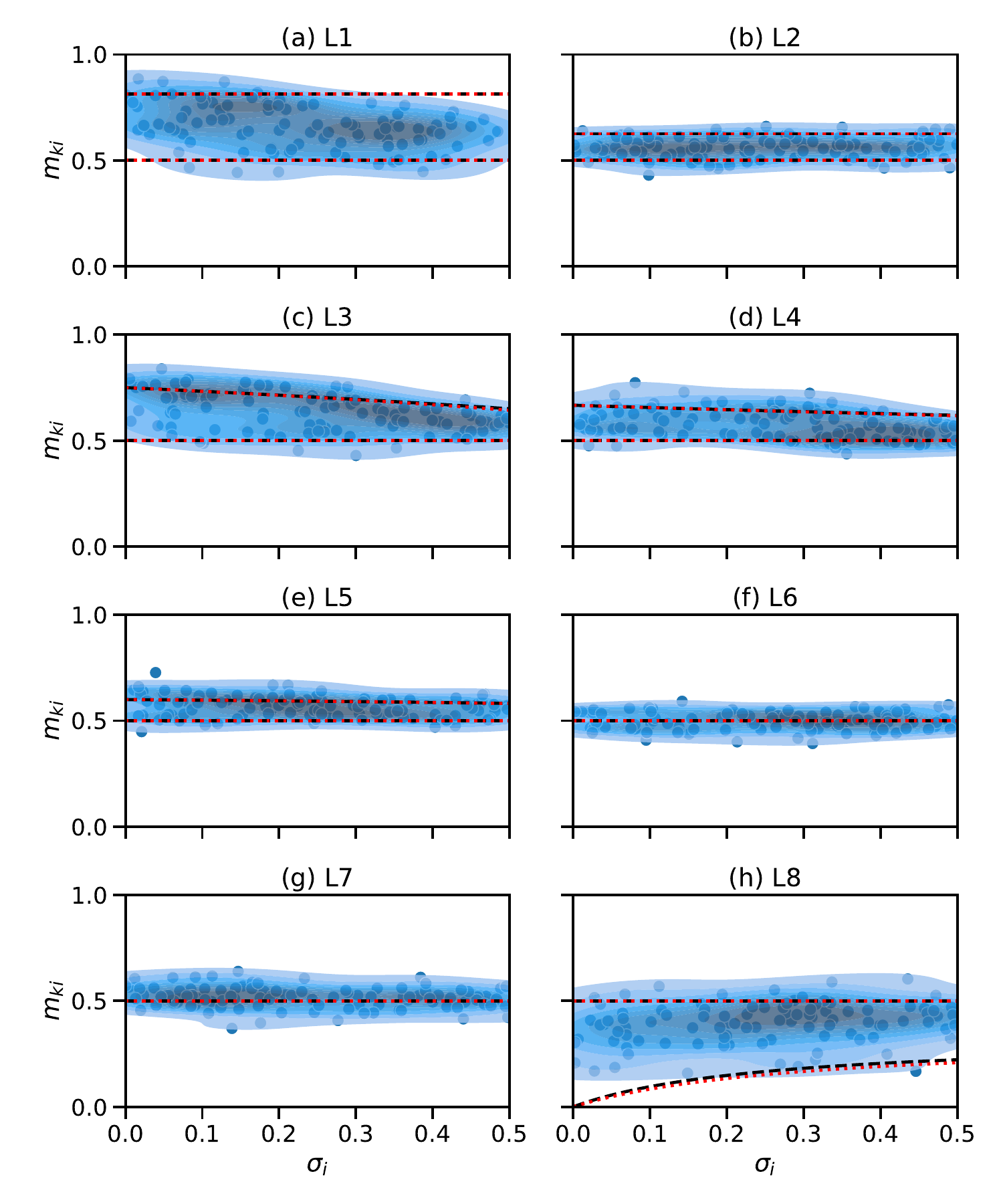}
\caption{$m_{ki}$ plotted against $\sigma_i$, instead of $\sigma_k$. We have used the same data as in Fig.~\ref{fig:out}. The black dashed lines show the cases of $N=2$, and the red dotted lines show the cases of $N=3$ in Table~\ref{tab:observer}.
}
\label{fig:in}
\end{figure}

As we have already seen, when assessment errors are caused by observers, that is, $\sigma_{ki}=\sigma_k$, the dominant factor determining $m_{ki}$ should be the probability of error of the individual who makes the assessment, i.e., $\sigma_k$, but the working hypothesis here is that $\sigma_i$ of the one being assessed can also affect the assessment in the long run.
Our analysis given above effectively corresponds to a one-body problem [see, e.g., the derivation of Eq.~\eqref{eq:one-body}], which can be obtained from Eq.~\eqref{eq:equation} by taking $N=1$. To estimate how others assess an individual with a finite error probability, we need $N=2$ at least, and we will apply the Newton method again to find an approximate solution. In a $d$-dimensional problem given by $f_1(x_1, x_2, \ldots, x_d) = f_2(x_1, x_2, \ldots, x_d) = \ldots = f_d(x_1, x_2, \ldots, x_d) = 0$, Eq.~\eqref{eq:Newton} is rewritten as
\begin{equation}
\begin{pmatrix}
x_1^\ast\\
x_2^\ast\\
\vdots\\
x_d^\ast
\end{pmatrix}
\approx
\begin{pmatrix}
\hat{x}_1\\
\hat{x}_2\\
\vdots\\
\hat{x}_d
\end{pmatrix}
-
\begin{pmatrix}
\frac{\partial f_1}{\partial x_1} & \frac{\partial f_1}{\partial x_2} & \ldots & \frac{\partial f_1}{\partial x_d}\\
\frac{\partial f_2}{\partial x_1} & \frac{\partial f_2}{\partial x_2} & \ldots & \frac{\partial f_2}{\partial x_d}\\
\vdots & \vdots & \ddots & \vdots\\
\frac{\partial f_d}{\partial x_1} & \frac{\partial f_d}{\partial x_2} & \ldots & \frac{\partial f_d}{\partial x_d}\\
\end{pmatrix}^{-1}
\begin{pmatrix}
f_1 (\hat{x}_1, \hat{x}_2, \ldots, \hat{x}_d)\\
f_2 (\hat{x}_1, \hat{x}_2, \ldots, \hat{x}_d)\\
\vdots\\
f_d (\hat{x}_1, \hat{x}_2, \ldots, \hat{x}_d)
\end{pmatrix},
\label{eq:high_dim}
\end{equation}
where the terms on the right-hand side are all evaluated at the trial solution $(\hat{x}_1, \hat{x}_2, \ldots, \hat{x}_d)$.
The resulting solution $\left\{ m_{ki}=m_{ki}^\ast \right\}$ generally depends on both $\sigma_k$ and $\sigma_i$, and we present the range of $m_{ki}$ for each norm as a function of $\sigma_i$ to remove the dependency on $\sigma_k$ in Table~\ref{tab:observer}. We have performed the analytic calculation only for $N=2$ and $3$, but the size dependence appears to be weak, and the predicted ranges are qualitatively consistent with the numerical data (Fig.~\ref{fig:in}).

\section{Discussion}

\begin{figure}
\centering
\includegraphics[width=0.5\linewidth]{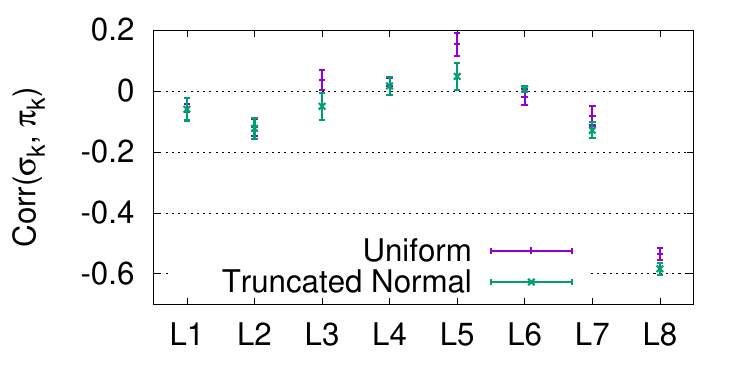}
\caption{Pearson correlation between an individual $k$'s payoff $\pi_k$ and the error probability $\sigma_k$, when the benefit of cooperation is $b=1$ whereas the cost is $c=1/2$.
The simulation details are the same as in Fig.~\ref{fig:out}. We have recorded a specific individual's payoff [Eq.~\eqref{eq:payoff}] at the end of each run and calculated the correlation coefficient from $10^2$ such data points. To estimate the mean and the standard error, we have repeated this procedure $10$ times. The error probabilities $\left\{\sigma_{k}\right\}$ are drawn from $\left[0,\frac{1}{2}\right)$ either uniformly  or by following the truncated normal distribution whose center and width are $0$ and $1/4$, respectively.}
\label{fig:corr}
\end{figure}

The important point of the above analysis is that the correlation of $m_{ki}$ with $\sigma_i$ is found to be weaker than the correlation with $\sigma_k$. For example, if we look at L8, the amount of donation that an individual $k$ gives to others increases with $\sigma_k$ [Fig.~\ref{fig:out}(h)], while the amount of cooperation that he or she receives is almost insensitive to $\sigma_k$ [Fig.~\ref{fig:in}(h)]. If we define an individual's payoff in the following way,
\begin{equation}
\pi_k \equiv \frac{1}{N} \sum_{i=1}^N \left[ b \beta\left(m_{ii}, m_{ik} \right) - c \beta\left(m_{kk}, m_{ki} \right) \right],
\label{eq:payoff}
\end{equation}
it is thus expected to decrease under the action of L8 as $\sigma_k$ grows. This prediction is verified by our numerical results (Fig.~\ref{fig:corr}). Among the leading eight, L8 exhibits the clearest signal of negative correlation between $\sigma_k$ and $\pi_k$. The more often one makes mistakes, the lower the payoff. The correct assessment is that others are bad [Fig.~\ref{fig:out}(h)], so the level of cooperation is low.
In contrast, even if L3 shows a higher level of cooperation~\cite{lee2022second,fujimoto2024leader}, Fig.~\ref{fig:corr} shows that it does not impose a heavy penalty for error-proneness.
According to Eq.~\eqref{eq:mkk_general},
the decisive factor for determining the sign of the correlation is $\alpha^\ast \equiv \alpha \left[ \frac{1}{2}, \beta\left(\frac{1}{2}, \frac{1}{2} \right), \frac{1}{2} \right]$, which makes sense in the continuous model where everyone can have a half-good state, but is also proved useful in the discrete model (Fig.~\ref{fig:out}), showing the power of the continuous model. If the social norm cannot support the half-good state by itself, that is, if $\alpha^\ast < 1/2$, a careful individual with low $\sigma_k$ will regard others as bad, securing his or her own payoff, although the average level of cooperation remains finite in the population.
Although the derivation in Eq.~\eqref{eq:sigma_k^n} uses the information on how the error probabilities are distributed in the population, Fig.~\ref{fig:corr} shows that L8 still has the most negative correlation even if the distribution is not uniform. We also observe the same trend for larger populations (not shown).

As for behavioral errors, by which one chooses defection although cooperation is intended and vice versa, we may expect that all norms in the leading eight will punish error-prone individuals because they have been designed to suppress behavioral mutants~\cite{ohtsuki2004should}, which can mimic behavioral errors. Our numerical calculations show that this is indeed the case, with two trivial exceptions having negligible correlations. One is L6, totally driven by entropy even with an arbitrarily small error probability~\cite{bae2024exact}, and the other is L8, under which everyone is eventually considered bad.

\section{Summary}

In summary, we have investigated how a social norm shapes society in the presence of heterogeneity in individual probabilities of assessment errors.
So far, the stability of a social norm has usually been analyzed in terms of error and mutation, but one of the theoretical challenges is to deal with these sources of randomness within a reasonable number of degrees of freedom. Concerning error, the traditional approach assumes that everyone is equally prone to it. When it comes to mutation, although it introduces heterogeneity in the population, most studies restrict themselves to the low mutation limit to simplify the problem (see, however, Ref.~\cite{vasconcelos2017stochastic} as an exception).

This work has extended the concept of stability by considering individual heterogeneity in assessment errors as a quenched disorder and has shown the possibility of analytic understanding.
Among the leading eight, L8 (Judging) strongly punishes those who do not carefully follow the norm, and it suggests the unique power of Judging when there exists heterogeneity among individuals.
The important factor turns out to be how the norm assesses the situation where everyone is half good, although it is a hypothetical point that makes full sense when the norm is interpolated between good and bad. It confirms the necessity and usefulness of the continuous model of indirect reciprocity.

\begin{acknowledgments}
We acknowledge support by Basic Science Research Program through the National Research Foundation of Korea (NRF) funded by the Ministry of Education (NRF-2020R1I1A2071670).
\end{acknowledgments}

\bibliography{apssamp}

\begin{thebibliography}{21}%
\makeatletter
\providecommand \@ifxundefined [1]{%
 \@ifx{#1\undefined}
}%
\providecommand \@ifnum [1]{%
 \ifnum #1\expandafter \@firstoftwo
 \else \expandafter \@secondoftwo
 \fi
}%
\providecommand \@ifx [1]{%
 \ifx #1\expandafter \@firstoftwo
 \else \expandafter \@secondoftwo
 \fi
}%
\providecommand \natexlab [1]{#1}%
\providecommand \enquote  [1]{``#1''}%
\providecommand \bibnamefont  [1]{#1}%
\providecommand \bibfnamefont [1]{#1}%
\providecommand \citenamefont [1]{#1}%
\providecommand \href@noop [0]{\@secondoftwo}%
\providecommand \href [0]{\begingroup \@sanitize@url \@href}%
\providecommand \@href[1]{\@@startlink{#1}\@@href}%
\providecommand \@@href[1]{\endgroup#1\@@endlink}%
\providecommand \@sanitize@url [0]{\catcode `\\12\catcode `\$12\catcode `\&12\catcode `\#12\catcode `\^12\catcode `\_12\catcode `\%12\relax}%
\providecommand \@@startlink[1]{}%
\providecommand \@@endlink[0]{}%
\providecommand \url  [0]{\begingroup\@sanitize@url \@url }%
\providecommand \@url [1]{\endgroup\@href {#1}{\urlprefix }}%
\providecommand \urlprefix  [0]{URL }%
\providecommand \Eprint [0]{\href }%
\providecommand \doibase [0]{http://dx.doi.org/}%
\providecommand \selectlanguage [0]{\@gobble}%
\providecommand \bibinfo  [0]{\@secondoftwo}%
\providecommand \bibfield  [0]{\@secondoftwo}%
\providecommand \translation [1]{[#1]}%
\providecommand \BibitemOpen [0]{}%
\providecommand \bibitemStop [0]{}%
\providecommand \bibitemNoStop [0]{.\EOS\space}%
\providecommand \EOS [0]{\spacefactor3000\relax}%
\providecommand \BibitemShut  [1]{\csname bibitem#1\endcsname}%
\let\auto@bib@innerbib\@empty
\bibitem [{\citenamefont {Nowak}\ and\ \citenamefont {Sigmund}(1998)}]{nowak1998evolution}%
  \BibitemOpen
  \bibfield  {author} {\bibinfo {author} {\bibfnamefont {M.~A.}\ \bibnamefont {Nowak}}\ and\ \bibinfo {author} {\bibfnamefont {K.}~\bibnamefont {Sigmund}},\ }\href@noop {} {\bibfield  {journal} {\bibinfo  {journal} {Nature}\ }\textbf {\bibinfo {volume} {393}},\ \bibinfo {pages} {573} (\bibinfo {year} {1998})}\BibitemShut {NoStop}%
\bibitem [{\citenamefont {Nowak}\ and\ \citenamefont {Sigmund}(2005)}]{nowak2005evolution}%
  \BibitemOpen
  \bibfield  {author} {\bibinfo {author} {\bibfnamefont {M.~A.}\ \bibnamefont {Nowak}}\ and\ \bibinfo {author} {\bibfnamefont {K.}~\bibnamefont {Sigmund}},\ }\href@noop {} {\bibfield  {journal} {\bibinfo  {journal} {Nature}\ }\textbf {\bibinfo {volume} {437}},\ \bibinfo {pages} {1291} (\bibinfo {year} {2005})}\BibitemShut {NoStop}%
\bibitem [{\citenamefont {Ohtsuki}\ and\ \citenamefont {Iwasa}(2004)}]{ohtsuki2004should}%
  \BibitemOpen
  \bibfield  {author} {\bibinfo {author} {\bibfnamefont {H.}~\bibnamefont {Ohtsuki}}\ and\ \bibinfo {author} {\bibfnamefont {Y.}~\bibnamefont {Iwasa}},\ }\href@noop {} {\bibfield  {journal} {\bibinfo  {journal} {J. Theor. Biol.}\ }\textbf {\bibinfo {volume} {231}},\ \bibinfo {pages} {107} (\bibinfo {year} {2004})}\BibitemShut {NoStop}%
\bibitem [{\citenamefont {Ohtsuki}\ and\ \citenamefont {Iwasa}(2006)}]{ohtsuki2006leading}%
  \BibitemOpen
  \bibfield  {author} {\bibinfo {author} {\bibfnamefont {H.}~\bibnamefont {Ohtsuki}}\ and\ \bibinfo {author} {\bibfnamefont {Y.}~\bibnamefont {Iwasa}},\ }\href@noop {} {\bibfield  {journal} {\bibinfo  {journal} {J. Theor. Biol.}\ }\textbf {\bibinfo {volume} {239}},\ \bibinfo {pages} {435} (\bibinfo {year} {2006})}\BibitemShut {NoStop}%
\bibitem [{\citenamefont {Ohtsuki}\ \emph {et~al.}(2009)\citenamefont {Ohtsuki}, \citenamefont {Iwasa},\ and\ \citenamefont {Nowak}}]{ohtsuki2009indirect}%
  \BibitemOpen
  \bibfield  {author} {\bibinfo {author} {\bibfnamefont {H.}~\bibnamefont {Ohtsuki}}, \bibinfo {author} {\bibfnamefont {Y.}~\bibnamefont {Iwasa}}, \ and\ \bibinfo {author} {\bibfnamefont {M.~A.}\ \bibnamefont {Nowak}},\ }\href@noop {} {\bibfield  {journal} {\bibinfo  {journal} {Nature}\ }\textbf {\bibinfo {volume} {457}},\ \bibinfo {pages} {79} (\bibinfo {year} {2009})}\BibitemShut {NoStop}%
\bibitem [{\citenamefont {Hilbe}\ \emph {et~al.}(2018)\citenamefont {Hilbe}, \citenamefont {Schmid}, \citenamefont {Tkadlec}, \citenamefont {Chatterjee},\ and\ \citenamefont {Nowak}}]{hilbe2018indirect}%
  \BibitemOpen
  \bibfield  {author} {\bibinfo {author} {\bibfnamefont {C.}~\bibnamefont {Hilbe}}, \bibinfo {author} {\bibfnamefont {L.}~\bibnamefont {Schmid}}, \bibinfo {author} {\bibfnamefont {J.}~\bibnamefont {Tkadlec}}, \bibinfo {author} {\bibfnamefont {K.}~\bibnamefont {Chatterjee}}, \ and\ \bibinfo {author} {\bibfnamefont {M.~A.}\ \bibnamefont {Nowak}},\ }\href@noop {} {\bibfield  {journal} {\bibinfo  {journal} {Proc. Natl. Acad. Sci. USA}\ }\textbf {\bibinfo {volume} {115}},\ \bibinfo {pages} {12241} (\bibinfo {year} {2018})}\BibitemShut {NoStop}%
\bibitem [{\citenamefont {Murase}\ and\ \citenamefont {Hilbe}(2024)}]{murase2024computational}%
  \BibitemOpen
  \bibfield  {author} {\bibinfo {author} {\bibfnamefont {Y.}~\bibnamefont {Murase}}\ and\ \bibinfo {author} {\bibfnamefont {C.}~\bibnamefont {Hilbe}},\ }\href@noop {} {\bibfield  {journal} {\bibinfo  {journal} {Proc. Natl. Acad. Sci. USA}\ }\textbf {\bibinfo {volume} {121}},\ \bibinfo {pages} {e2406885121} (\bibinfo {year} {2024})}\BibitemShut {NoStop}%
\bibitem [{\citenamefont {Lee}\ \emph {et~al.}(2022)\citenamefont {Lee}, \citenamefont {Murase},\ and\ \citenamefont {Baek}}]{lee2022second}%
  \BibitemOpen
  \bibfield  {author} {\bibinfo {author} {\bibfnamefont {S.}~\bibnamefont {Lee}}, \bibinfo {author} {\bibfnamefont {Y.}~\bibnamefont {Murase}}, \ and\ \bibinfo {author} {\bibfnamefont {S.~K.}\ \bibnamefont {Baek}},\ }\href@noop {} {\bibfield  {journal} {\bibinfo  {journal} {J. Theor. Biol.}\ }\textbf {\bibinfo {volume} {548}},\ \bibinfo {pages} {111202} (\bibinfo {year} {2022})}\BibitemShut {NoStop}%
\bibitem [{\citenamefont {Bae}\ \emph {et~al.}(2024)\citenamefont {Bae}, \citenamefont {Shimada},\ and\ \citenamefont {Baek}}]{bae2024exact}%
  \BibitemOpen
  \bibfield  {author} {\bibinfo {author} {\bibfnamefont {M.}~\bibnamefont {Bae}}, \bibinfo {author} {\bibfnamefont {T.}~\bibnamefont {Shimada}}, \ and\ \bibinfo {author} {\bibfnamefont {S.~K.}\ \bibnamefont {Baek}},\ }\href@noop {} {\bibfield  {journal} {\bibinfo  {journal} {Phys. Rev. E}\ }\textbf {\bibinfo {volume} {110}},\ \bibinfo {pages} {L052301} (\bibinfo {year} {2024})}\BibitemShut {NoStop}%
\bibitem [{\citenamefont {Easley}\ and\ \citenamefont {Kleinberg}(2010)}]{easley2010networks}%
  \BibitemOpen
  \bibfield  {author} {\bibinfo {author} {\bibfnamefont {D.}~\bibnamefont {Easley}}\ and\ \bibinfo {author} {\bibfnamefont {J.}~\bibnamefont {Kleinberg}},\ }\enquote {\bibinfo {title} {A weaker form of structural balance},}\ in\ \href@noop {} {\emph {\bibinfo {booktitle} {Networks, Crowds, and Markets: Reasoning about a Highly Connected World}}}\ (\bibinfo  {publisher} {Cambridge University Press},\ \bibinfo {address} {Cambridge},\ \bibinfo {year} {2010})\ Chap.~\bibinfo {chapter} {5}, pp.\ \bibinfo {pages} {115--118}\BibitemShut {NoStop}%
\bibitem [{\citenamefont {Bae}\ and\ \citenamefont {Baek}(2025)}]{bae2025indirect}%
  \BibitemOpen
  \bibfield  {author} {\bibinfo {author} {\bibfnamefont {M.}~\bibnamefont {Bae}}\ and\ \bibinfo {author} {\bibfnamefont {S.~K.}\ \bibnamefont {Baek}},\ }\href@noop {} {\enquote {\bibinfo {title} {Indirect reciprocity as a dynamics for weak balance},}\ }\bibinfo {howpublished} {arXiv:2501.05824} (\bibinfo {year} {2025})\BibitemShut {NoStop}%
\bibitem [{\citenamefont {Fujimoto}\ and\ \citenamefont {Ohtsuki}(2024)}]{fujimoto2024leader}%
  \BibitemOpen
  \bibfield  {author} {\bibinfo {author} {\bibfnamefont {Y.}~\bibnamefont {Fujimoto}}\ and\ \bibinfo {author} {\bibfnamefont {H.}~\bibnamefont {Ohtsuki}},\ }\href@noop {} {\bibfield  {journal} {\bibinfo  {journal} {PRX Life}\ }\textbf {\bibinfo {volume} {2}},\ \bibinfo {pages} {023009} (\bibinfo {year} {2024})}\BibitemShut {NoStop}%
\bibitem [{\citenamefont {Leskovec}\ \emph {et~al.}(2010)\citenamefont {Leskovec}, \citenamefont {Huttenlocher},\ and\ \citenamefont {Kleinberg}}]{leskovec2010signed}%
  \BibitemOpen
  \bibfield  {author} {\bibinfo {author} {\bibfnamefont {J.}~\bibnamefont {Leskovec}}, \bibinfo {author} {\bibfnamefont {D.}~\bibnamefont {Huttenlocher}}, \ and\ \bibinfo {author} {\bibfnamefont {J.}~\bibnamefont {Kleinberg}},\ }in\ \href@noop {} {\emph {\bibinfo {booktitle} {Proc. SIGCHI Conf. Hum. Factor Comput. Syst.}}}\ (\bibinfo  {publisher} {Association for Computing Machinery},\ \bibinfo {address} {New York, NY},\ \bibinfo {year} {2010})\ pp.\ \bibinfo {pages} {1361--1370}\BibitemShut {NoStop}%
\bibitem [{\citenamefont {Szell}\ \emph {et~al.}(2010)\citenamefont {Szell}, \citenamefont {Lambiotte},\ and\ \citenamefont {Thurner}}]{szell2010multirelational}%
  \BibitemOpen
  \bibfield  {author} {\bibinfo {author} {\bibfnamefont {M.}~\bibnamefont {Szell}}, \bibinfo {author} {\bibfnamefont {R.}~\bibnamefont {Lambiotte}}, \ and\ \bibinfo {author} {\bibfnamefont {S.}~\bibnamefont {Thurner}},\ }\href@noop {} {\bibfield  {journal} {\bibinfo  {journal} {Proc. Natl. Acad. Sci. USA}\ }\textbf {\bibinfo {volume} {107}},\ \bibinfo {pages} {13636} (\bibinfo {year} {2010})}\BibitemShut {NoStop}%
\bibitem [{\citenamefont {Lee}\ \emph {et~al.}(2021)\citenamefont {Lee}, \citenamefont {Murase},\ and\ \citenamefont {Baek}}]{lee2021local}%
  \BibitemOpen
  \bibfield  {author} {\bibinfo {author} {\bibfnamefont {S.}~\bibnamefont {Lee}}, \bibinfo {author} {\bibfnamefont {Y.}~\bibnamefont {Murase}}, \ and\ \bibinfo {author} {\bibfnamefont {S.~K.}\ \bibnamefont {Baek}},\ }\href@noop {} {\bibfield  {journal} {\bibinfo  {journal} {Sci. Rep.}\ }\textbf {\bibinfo {volume} {11}},\ \bibinfo {pages} {14225} (\bibinfo {year} {2021})}\BibitemShut {NoStop}%
\bibitem [{\citenamefont {Mun}\ and\ \citenamefont {Baek}(2024)}]{mun2023second}%
  \BibitemOpen
  \bibfield  {author} {\bibinfo {author} {\bibfnamefont {Y.}~\bibnamefont {Mun}}\ and\ \bibinfo {author} {\bibfnamefont {S.~K.}\ \bibnamefont {Baek}},\ }\href@noop {} {\bibfield  {journal} {\bibinfo  {journal} {Eur. Phys. J. Spec. Top.}\ }\textbf {\bibinfo {volume} {233}},\ \bibinfo {pages} {1251} (\bibinfo {year} {2024})}\BibitemShut {NoStop}%
\bibitem [{\citenamefont {Mun}\ \emph {et~al.}(2024)\citenamefont {Mun}, \citenamefont {Le},\ and\ \citenamefont {Baek}}]{mun2024making}%
  \BibitemOpen
  \bibfield  {author} {\bibinfo {author} {\bibfnamefont {Y.}~\bibnamefont {Mun}}, \bibinfo {author} {\bibfnamefont {Q.~A.}\ \bibnamefont {Le}}, \ and\ \bibinfo {author} {\bibfnamefont {S.~K.}\ \bibnamefont {Baek}},\ }\href@noop {} {\bibfield  {journal} {\bibinfo  {journal} {J. Korean Phys. Soc.}\ }\textbf {\bibinfo {volume} {85}},\ \bibinfo {pages} {969} (\bibinfo {year} {2024})}\BibitemShut {NoStop}%
\bibitem [{\citenamefont {Baek}\ \emph {et~al.}(2017)\citenamefont {Baek}, \citenamefont {Do~Yi},\ and\ \citenamefont {Jeong}}]{baek2017duality}%
  \BibitemOpen
  \bibfield  {author} {\bibinfo {author} {\bibfnamefont {S.~K.}\ \bibnamefont {Baek}}, \bibinfo {author} {\bibfnamefont {S.}~\bibnamefont {Do~Yi}}, \ and\ \bibinfo {author} {\bibfnamefont {H.-C.}\ \bibnamefont {Jeong}},\ }\href@noop {} {\bibfield  {journal} {\bibinfo  {journal} {J. Theor. Biol.}\ }\textbf {\bibinfo {volume} {430}},\ \bibinfo {pages} {215} (\bibinfo {year} {2017})}\BibitemShut {NoStop}%
\bibitem [{\citenamefont {Waskom}(2021)}]{Waskom2021}%
  \BibitemOpen
  \bibfield  {author} {\bibinfo {author} {\bibfnamefont {M.~L.}\ \bibnamefont {Waskom}},\ }\href {\doibase 10.21105/joss.03021} {\bibfield  {journal} {\bibinfo  {journal} {J. Open Source Softw.}\ }\textbf {\bibinfo {volume} {6}},\ \bibinfo {pages} {3021} (\bibinfo {year} {2021})}\BibitemShut {NoStop}%
\bibitem [{Mat()}]{Mathematica}%
  \BibitemOpen
  \href@noop {} {\enquote {\bibinfo {title} {Mathematica, {V}ersion 10.0},}\ }\bibinfo {note} {({Wolfram Research, Inc.}, {C}hampaign, IL, 2014)}\BibitemShut {NoStop}%
\bibitem [{\citenamefont {Vasconcelos}\ \emph {et~al.}(2017)\citenamefont {Vasconcelos}, \citenamefont {Santos}, \citenamefont {Santos},\ and\ \citenamefont {Pacheco}}]{vasconcelos2017stochastic}%
  \BibitemOpen
  \bibfield  {author} {\bibinfo {author} {\bibfnamefont {V.~V.}\ \bibnamefont {Vasconcelos}}, \bibinfo {author} {\bibfnamefont {F.~P.}\ \bibnamefont {Santos}}, \bibinfo {author} {\bibfnamefont {F.~C.}\ \bibnamefont {Santos}}, \ and\ \bibinfo {author} {\bibfnamefont {J.~M.}\ \bibnamefont {Pacheco}},\ }\href@noop {} {\bibfield  {journal} {\bibinfo  {journal} {Phys. Rev. Lett.}\ }\textbf {\bibinfo {volume} {118}},\ \bibinfo {pages} {058301} (\bibinfo {year} {2017})}\BibitemShut {NoStop}%
\end{thebibliography}%

\end{document}